# Constructing "Wavefunctions" for One-Body and Two-Body Gravitational Orbits in Classical Mechanics

*Jixin Chen**

*Nanoscale & Quantum Phenomena Institute, Department of Chemistry and Biochemistry, Ohio University, Athens Ohio 45701*

chenj@ohio.edu

## Abstract
The circular orbits and elliptical orbits of moving objects in a gravitational field are essential information in astronomy. There have been many methods developed in the literature and textbooks to describe these orbits. In this report, I propose to use the *vis-viva* equation to construct a complex function to store the state of a moving object in elliptical orbits such that one can calculate its near future numerically. This state function is constructed by splitting its momentum into real and imaginary parts with one perpendicular to the radius of the mass center and the other parallel. The wavefunctions of electrons of hydrogen atoms in quantum mechanics inspire this idea. The equations are derived for one-body problems. Two-body problems can be constructed with subsets of one-body problems with the same center of mass, but different effective mass pinned there, significantly different from existing methods and providing the same results.

## Keywords
Classical mechanics, elliptic orbit, complex state functions, planet and satellite.

## Introduction
Calculating the elliptic orbits of moving objects in a star system under gravitational forces is important to humankind with the greatest breakthrough many of us may agree to attribute to Isaac Newton's classical mechanics. The gravitational system is a similar condition to the electrons orbiting atomic cores sharing the same mathematical equation in forces being both dependent on one over square the distance between the objects. In the atomic system, the motions of electrons are described by wavefunctions proposed by Erwin Schrödinger which are sets of complex numbers. I was wondering if we can construct a "wavefunction" for classical mechanics, particularly the motions of objects with gravitational forces with elliptical orbits, which I am giving a try in this report, and you can see it might have some advantages of doing so.

## Results and discussion
### Orbits of one-body problems

The classical mechanics of planet orbits such as the Earth orbiting the Sun or satellites orbiting the Earth have been solved analytically e.g. with the *vis-viva* equation as reported in textbooks and literature.[1–6] Due to the Sun's massive mass relative to that of the Earth, this problem can be approximated as a one-body problem. The center of mass is very close to the center of the large object thus we can assume the large object does not move.

The small object (mass *m*) has potential energy due to the gravitational pull from the large object (mass *M*) assuming obeying Newton's law of gravity when setting the potential energy at infinity to be zero as the reference point:

$$E_V = \int_{\infty}^{r} \left(\frac{Gm_1m_2}{x^2}\right) dx = -\frac{GMm}{r} \quad (1)$$

where *G* is the gravitational constant and r is the distance between the centers of these two objects. The kinetic energy of the small object is:

$$E_k = \frac{1}{2}mv^2 \quad (2)$$

The total energy of the small object at any given condition at distance *r* is the sum of the potential energy and its kinetic energy:

$$E_{\text{tot}} = E_V + E_k \quad (3)$$

If $E_{\text{tot}} > 0$, there is a range of angles of *v* for the object to hit the surface of the larger object which is a sphere in our example, and another range to be parabolic or hyperbolic escape orbits. If $E_{\text{tot}} < 0$, the small object is caught by the gravitational field of the larger object whose orbit will be determined by the angle of *v* among which we are interested in the circular orbit and elliptic orbits.

Under a special condition when the centrifugal force balances the gravitational force in the opposite direction,

$$\frac{mv^2}{r} = \frac{GMm}{r^2} \quad (4)$$

with a symmetry argument, we can see that there is only one angle and value of *v* for a circular orbit when half of its potential energy from infinite far is lost and the angle of the velocity is perpendicular to the radius of the orbit:

$$E_V = -\frac{GMm}{r} \;;\; E_k = \frac{GMm}{2r} \;;\; E_{tot} = -\frac{GMm}{2r} \quad (5)$$

We can compute the equilibrium radius ($R_E$) of this circular orbit to any object with a ***conserved energy*** $E_{\text{tot}} < 0$, regardless of its current position:

$$R_E = -\frac{GMm}{2E_{\text{tot}}} \quad (6)$$

Under the special case when a stable circular orbit is formed, the angular momentum of the small object is,

$$L = \sqrt{GMm^2 R_E} \qquad (7)$$

This is the maximum angular momentum of the small object with total energy $E_{tot}$ because the velocity at $r = R_E$ is perpendicular to the radius, the line connecting the small and the large object, $v_\perp = v$ (**Fig. 1A**). All the rest cases of $E_{tot} < 0$, the small object follows an elliptic orbit because at $R_E$ when the values of the velocity are the same as their circular sibling, there is an angle deviate from that of the circular orbit.

We can set a parameter $0 <= a <= 1$ to describe the angular momentum of the small object, such that the angular momentum when it passes $R_E$ is:

$$L_a(@R_E) = mv_\perp R_E = \sqrt{aGMm^2 R_E} \qquad (8)$$

When $a = 1$, the object follows the circular orbit; when $0 < a < 1$, the object follows elliptic orbits; and when a = 0 which should not exist, the small object free falls towards the center of the large object. This factor aligns the opposite trend with the eccentricity of the ellipse that has been widely used in the literature and textbooks to describe this elliptical orbit problem.[1] Compared to eccentricity, the angular momentum factor is more consistent with the idea in quantum mechanics that the angular momentum is quantized.

We have learned from Kepler's 2nd law and Newton's laws of motion that the small object **conserves this angular momentum**, i.e. at any possible distance,

$$mv_\perp r = \sqrt{aGMm^2 R_E} \qquad (9)$$

At any given moment, if we assign the angles of the elliptic orbits with respect to the radius $r$ as θ (**Fig. 1A**), we can see that using the results from the *vis-viva* equation,

$$E_k = \frac{1}{2}mv^2 = E_{tot} - E_V = -\frac{GMm}{2R_E} + \frac{GMm}{r} \qquad (10)$$

$$v = \sqrt{-\frac{GM}{R_E} + \frac{2GM}{r}} \qquad (11)$$

Thus, we conclude that at radius $r$ with the angular momentum factor $\sqrt{a}$ and total energy $E_{tot}$,

$$v_\perp = \frac{\sqrt{aGMR_E}}{r} \qquad (12)$$

$$v_\parallel = \sqrt{-\frac{GM}{R_E} + \frac{2GM}{r} - \frac{aGMR_E}{r^2}} \qquad (13)$$

$$\cos\theta = \frac{v_\perp}{v} = \frac{\sqrt{-\frac{r^2}{R_E}+2r}}{\sqrt{aR_E}} = \sqrt{-\frac{r^2}{aR_E^2} + \frac{2r}{aR_E}} \qquad (14)$$

Thus, use the large object as the reference of frame, if we were able to measure the small object at any given time in space given its distance to the center of the large object, the mass of the large object, the velocity of the small object, and the angle with respect to the radius which defines the ecliptic plane, we can predict its entire orbit from this point of time into the future.

We can combine the two parts of momentum into a single complex number to describe the state of the small object which naturally build in the angle of the orbits respected to the concentered rings:

$$\psi(r) = mv_\perp + imv_\parallel = m\frac{\sqrt{aGMR_E}}{r} \pm im\sqrt{-\frac{GM}{R_E} + \frac{2GM}{r} - \frac{aGMR_E}{r^2}} \quad (15)$$

We can see that this wavefunction is a solution to $\frac{1}{2m}\psi\psi^* = E_k = E_{tot} - E_V$ for the orbits with conservation of angular momentum and conservation of total energy as the boundary conditions.

Both Newton's 2$^{nd}$ law and **Equation 15** can be used to numerically simulate the motion (**Fig. 1B**, **1C**). However, when using constant time step and the correlation $\Delta\theta = \frac{v_\perp \Delta t}{r}$ to calculate the rotation angles, the propagated error is larger than Newton's method because the step size near perigee is large. This reduced accuracy near perigee is general for both methods in simulating elliptical orbits with large eccentricity. The error is significantly reduced if take constant rotation angle step instead of constant time step in the simulation, within a range that the sampling density at apogee is not too small. In addition, the sign in **Equation 15** is tuned manually for the two sides of the long semiaxis now which is somewhat inconvenient. When leaving apogee, the next step radius is approximated to,

$$r_2 = r_1 \cos(\Delta\theta) - v_\parallel \Delta t \quad (16)$$

When leaving perigee, the next step radius is approximated to,

$$r_2 = \frac{r_1}{\cos(\Delta\theta)} + v_\parallel \Delta t \quad (17)$$

Optimizing these two approximation equations could give a better simulation accuracy.

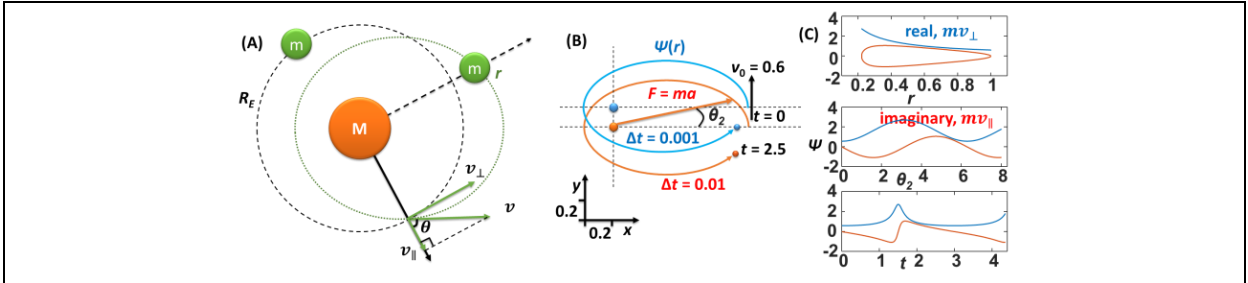

**Fig. 1.** (A) Scheme of gravitational circular orbit and elliptical orbit of two small objects (m) orbiting a large orbit with massive mass (M). The two small objects have the same total energy but different angular momenta. (B) The right figure compares two example numerical simulations using Newton's 2$^{rd}$ law and **Equation 15** with the same initial conditions. Two figures are shifted in the y axis for comparison. Set *GM* = *m* = 1 and $E_{Tot}$ = -0.82. The *x-y* plane is set to be an ecliptic plane so there is no motion in the *z*-axis in these simulations and convergence is not attempted either. (C) The wavefunction vs *r*, θ, and *t*. Please see supporting information for the source codes in MATLAB for both simulations and a movie.

## Orbits of two-body problems

For the two-body problem, a typical theory will set the center of mass (CoM) stationary over space and time as the reference point (**Fig. 2A**),[1,6–8] where

$$m_1 x_1 = m_2 x_2 \;;\; x_1 + x_2 = l \quad (18)$$

$$x_1 = \frac{m_2 l}{m_1+m_2} \quad ; \quad x_2 = \frac{m_1 l}{m_1+m_2} \qquad (19)$$

Thus, according to Newton's 3rd law, the conservation of momentum gives the correlation between the motion of the two bodies (**Fig. 2B**):

$$m_2 \vec{v_2} = -m_1 \vec{v_1} \qquad (20)$$

In both Newtonian mechanics and Lagrangian mechanics, the reduced mass is,

$$\mu = \frac{m_1 m_2}{m_1+m_2} \qquad (21)$$

which is used to calculate the relative motion between the two bodies. However, it is difficult to construct a rotation system with respect to the CoM using this reduced mass. We shall split the two bodies and create separated effective masses for the two subsystems in order to use the results of the one-body system we have constructed in the last section (**Fig. 2C**).

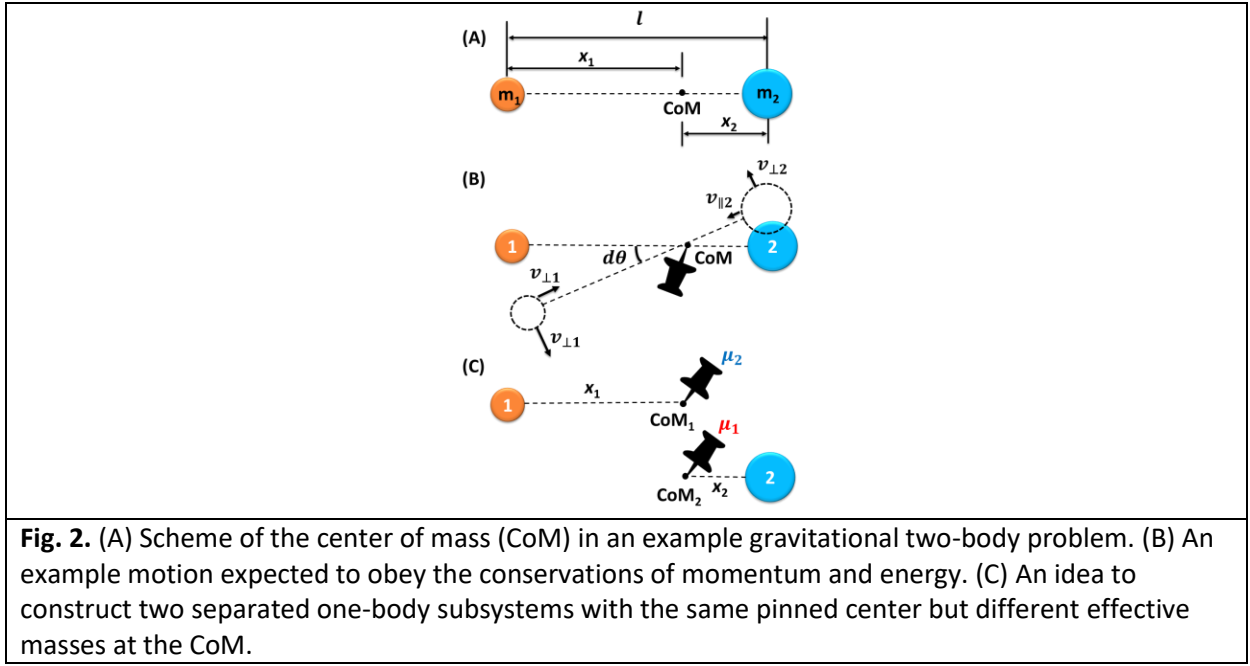

**Fig. 2.** (A) Scheme of the center of mass (CoM) in an example gravitational two-body problem. (B) An example motion expected to obey the conservations of momentum and energy. (C) An idea to construct two separated one-body subsystems with the same pinned center but different effective masses at the CoM.

For body $m_1$, the effective mass pinned at the CoM is,

$$\mu_2 = \frac{m_2^3}{(m_1+m_2)^2} \qquad (22)$$

such that the force $m_1$ feels from this effective mass pinned at CoM equals the force it feels from $m_2$ all the time due to synchronized motion. And for mass $m_2$, the effective mass pinned at CoM is,

$$\mu_1 = \frac{m_1^3}{(m_1+m_2)^2} \qquad (23)$$

We split the two-body problem into two one-body problems with the same pinned center but different effective masses. Setting the potential energy at infinite far to be zero, the potential energy is also split to be integration of two forces,

$$E_{V\_tot}(\infty \to l) = E_{V1} + E_{V2} = -\frac{Gm_1\mu_2}{x_1} - \frac{Gm_2\mu_1}{x_2} = -\frac{Gm_1m_2}{l} \qquad (24)$$

which is consistent with integrating the single force over the distance (**Equation 1**).

Thus, with an initial angular momentum, the rotational motion between the two bodies can be split into two different one-body rotations with respect to the center of mass at distance $r$, e.g. for $m_1$,

$$E_1 = E_{V1} + E_{k1} = -\frac{Gm_1\mu_2}{r_1} + \frac{1}{2}m_1v_1^2 \qquad (25)$$

$$R_{E1} = -\frac{Gm_1\mu_2}{2E_1} \qquad (26)$$

$$L_{a1}(@r_1) = m_1 v_{\perp 1} r_1 = \sqrt{a_1 G \mu_2 m_1^2 R_{E1}} \qquad (27)$$

$$\psi_1(r) = m_1 v_{\perp 1} + i m_1 v_{\parallel 1} = m_1 \frac{\sqrt{a_1 G \mu_2 R_{E1}}}{r_1} \pm i m_1 \sqrt{-\frac{G\mu_2}{R_{E1}} + \frac{2G\mu_2}{r_1} - \frac{a_1 G \mu_2 R_{E1}}{r_1^2}} \qquad (28)$$

where $R_E$ is the equivalent circular orbit radius and $0 <= a_1 <= 1$ is its angular momentum factor. The same set of equations can be derived for $m_2$, just switch the subscription 1 and 2. We can also derive from **Equations 20**, **26**, and **27** that $a_2 = a_1$, which makes sense since both orbits have the same shape.

The conservation of energy can be expressed as a Hamiltonian,

$$\frac{1}{2m_1}\psi_1\psi_1^* + \frac{1}{2m_2}\psi_2\psi_2^* = E_{k1} + E_{k2} = E_{tot} - E_V \qquad (29)$$

Splitting a two-body problem into two separated one-body problems makes it easy to judge if the orbit is circular ($a = 1$), ellipse ($a < 1$), or nonstable ($E_{tot} >= 0$). **Fig. 3** shows two simulations using Newton's laws of motion and the wavefunction method demonstrating consistency between these two methods. From Einstein's point of view, considering it takes time for one subset to update the information to the CoM with a limit at the speed of light, the two subsets will have CoMs at different locations and the rotation could be affected to induce precessions, while there is no such problem for a one-body system.

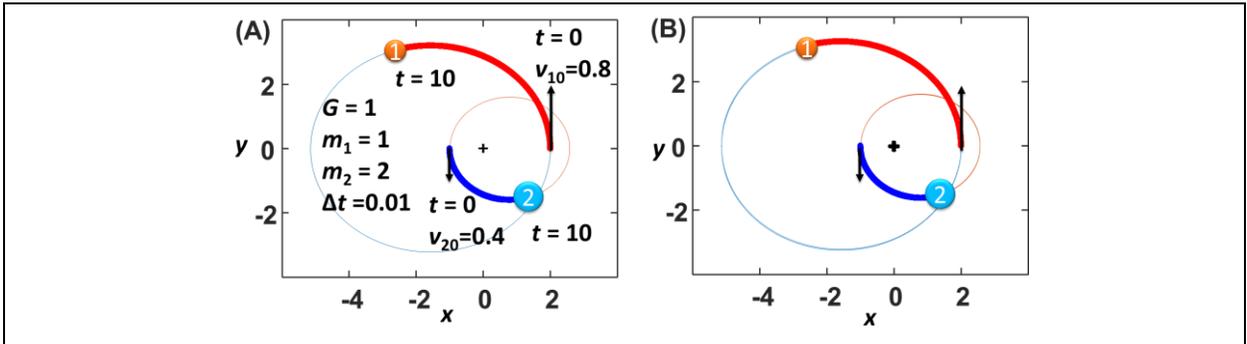

**Fig. 3.** Comparison of simulations with (A) Newton's laws of motion, and (B) wavefunction method with the same even time steps and initial conditions for an example two-body problem. The two simulations show the same results in both space and time with a bigger error observed for (B) with the same step time. Please see supporting information for the source codes for the two simulations and the movie.

Because all real gravitational systems are at least two-body problems, this method of splitting the system into subsets could be beneficial. The effective masses at the CoM in the two-body problem remain unchanged over time. **Many-body problems** cannot be solved by simply putting an effective mass at the CoM for each particle because the other particles have an overall drag force that is perpendicular to its radius of motion to the CoM. In addition, although the overall angular momentum remains the same for the system but the distribution among particles changes over time. It is challenging to construct a wavefunction without calculating the overall force and its direction.

## Conclusion

In summary, constructing a wavefunction to a classical mechanic problem of elliptical orbits in a gravitational system yields surprising simplicity of solving the problem. This could be a new way to calculate complicated planets and satellites orbits in a different angle of view from the established methods. The basic idea is to construct a one-body problem with a pinned center and a force between the center and the object following the gravitational laws. Then for two-body and many-body problems, subsets of one-body problems can be constructed with different pinned centers that have different effective masses even if the centers overlap in different subsets, e.g. the center of mass which is often chosen as the reference frame. I believe this is a new idea that has not been reported before to the best of knowledge of mine and I hope you agree that it is interesting to think this way.

## Author information

*Corresponding author. Email: chenj@ohio.edu

ORCID: 0000-0001-7381-0918

## Supporting information available

MATLAB source code available at https://github.com/nkchenjx/TwoBodyProblem

## Acknowledgment

Chen thanks his family for their support.

## Author Declarations

The authors have no conflicts of interest to disclose.

## Data Availability

All data is included in the main text and the supporting information.